\documentstyle[fleqn,12pt]{article}

\newif\ifPostScript
\PostScriptfalse

\title{Critical Exponents for the Metal--Insulator Transition in Disordered
Systems}
\author{Angus MacKinnon\\
Blackett Lab., Imperial College, London SW7 2BZ}
\newcommand{\shorttitle}[1]
   {\vfill
    \noindent \rm Short title: {\sl #1}\par
    \medskip}
\newcommand{\pacs}[1]
   {\noindent \rm PACS number(s): #1\par
    \medskip}
\newcommand{\jnl}[1]
   {\noindent \rm Submitted to: {\sl #1}\par
    \medskip}

\def\be{\begin{equation}}
\def\ee#1{\label{#1}\end{equation}}
\def\bes{\begin{subequations}\begin{eqnarray}}
\def\ees#1{\end{eqnarray}\label{#1}\end{subequations}}
\def\ba{\begin{array}}
\def\ea{\end{array}}

\let\dunder=\d
\renewcommand{\d}{\ifmmode{\mathord{\rm d}}\else\dunder\fi}
\let\dotlessi=\i
\renewcommand{\i}{\ifmmode{\mathord{\rm i\/}}\else\dotlessi\fi}

\newcommand{\half}{{\mathchoice{{\displaystyle{\frac12}}}%
{{\textstyle{1/2}}}{{\scriptstyle{1/2}}}{{\scriptscriptstyle{1/2}}}}}

\ifPostScript
   \font\sixbdi=ptmbi at 6pt
   \font\egtbdi=ptmbi at 8pt
   \font\twlbdi=ptmbi at 12pt
   \font\sixsans=phvb at 6pt
   \font\egtsans=phvb at 8pt
   \font\twlsans=phvb at 12pt
\else
   \font\sixbdi=cmmib10 at 6pt
   \font\egtbdi=cmmib10 at 8pt
   \font\twlbdi=cmmib10 scaled \magstep1
   \font\sixsans=cmssbx10 at 6pt
   \font\egtsans=cmssbx10 at 8pt
   \font\twlsans=cmssbx10 scaled \magstep1
\fi

\def\bi#1{\mathchoice{\hbox{\twlbdi #1}}{\hbox{\twlbdi #1}}{\hbox{\egtbdi
#1}}{\hbox{\sixbdi #1}}}
\def\bss#1{\mathchoice{\hbox{\twlsans #1}}{\hbox{\twlsans #1}}{\hbox{\egtsans
#1}}{\hbox{\sixsans #1}}}

\catcode`@=11
\def\@cite#1#2{\ifdim\lastskip>0pt\relax\else\space\fi(#1\if@tempswa , #2\fi)}
\def\@biblabel#1{}

\newlength{\bibhang}
\def\bibtitle{References}
\def\bibheadtitle{REFERENCES}

\@ifundefined{chapter}{\def\thebibliography#1{\section*{\bibtitle\@mkboth
  {\bibheadtitle}{\bibheadtitle}}
  \addcontentsline{toc}{section}{\bibtitle}\list            
  {\relax}{\setlength{\labelsep}{0em}                       
        \setlength{\itemindent}{-\bibhang}                  
        \setlength{\leftmargin}{\bibhang}}
    \def\newblock{\hskip .11em plus .33em minus .07em}
    \sloppy\clubpenalty4000\widowpenalty4000
    \sfcode`\.=1000\relax}}%
{\def\thebibliography#1{\chapter*{\bibtitle\@mkboth           
  {\bibheadtitle}{\bibheadtitle}}                             
  \addcontentsline{toc}{chapter}{\bibtitle}\list              
  {\relax}{\setlength{\labelsep}{0em}
        \setlength{\itemindent}{-\bibhang}
        \setlength{\leftmargin}{\bibhang}}
    \def\newblock{\hskip .11em plus .33em minus .07em}
    \sloppy\clubpenalty4000\widowpenalty4000
    \sfcode`\.=1000\relax}}
\def\@citex[#1]#2{\if@filesw\immediate\write\@auxout{\string\citation{#2}}\fi
  \def\@citea{}\@cite{\@for\@citeb:=#2\do
    {\@citea\def\@citea{; }\@ifundefined
       {b@\@citeb}{{\bf ?}\@warning
       {Citation `\@citeb' on page \thepage \space undefined}}%
{\csname b@\@citeb\endcsname}}}{#1}}
\let\@internalcite\cite
\def\cite{\def\citename##1{\citen@me{##1}}\def\@ldcite{\relax}\@internalcite}
\def\shortcite{\def\citename##1{}\@internalcite}
\def\citen@me#1{\def\n@wcite{#1}%
\ifx\n@wcite\@ldcite\relax\else\def\@ldcite{#1}#1\fi}

\newif\if@defeqnsw \@defeqnswtrue

\def\eqnarray{\stepcounter{equation}\let\@currentlabel=\theequation
\if@defeqnsw\global\@eqnswtrue\else\global\@eqnswfalse\fi
\global\@eqnswtrue
\tabskip\@centering\let\\=\@eqncr
$$\halign to \displaywidth\bgroup\hfil\global\@eqcnt\z@
  $\displaystyle\tabskip\z@{##}$&\global\@eqcnt\@ne
  \hfil$\displaystyle{{}##{}}$\hfil
  &\global\@eqcnt\tw@ $\displaystyle{##}$\hfil
  \tabskip\@centering&\llap{##}\tabskip\z@\cr}

\def\yesnumber{\global\@eqnswtrue}

\def\@@eqncr{\let\@tempa\relax\global\advance\@eqcnt by \@ne
    \ifcase\@eqcnt \def\@tempa{& & & &}\or \def\@tempa{& & &}\or
     \def\@tempa{& &}\or \def\@tempa{&}\else\fi
     \@tempa \if@eqnsw\@eqnnum\stepcounter{equation}\fi
     \if@defeqnsw\global\@eqnswtrue\else\global\@eqnswfalse\fi
     \global\@eqcnt\z@\cr}

\def\@eqnacr{{\ifnum0=`}\fi\@ifstar{\@yeqnacr}{\@yeqnacr}}

\def\@yeqnacr{\@ifnextchar [{\@xeqnacr}{\@xeqnacr[\z@]}}

\def\@xeqnacr[#1]{\ifnum0=`{\fi}\cr \noalign{\vskip\jot\vskip #1\relax}}

\def\eqalign{\null\,\vcenter\bgroup\openup1\jot \m@th \let\\=\@eqnacr
\ialign\bgroup\strut
\hfil$\displaystyle{##}$&$\displaystyle{{}##}$\hfil\crcr}
\def\endeqalign{\crcr\egroup\egroup\,}

\def\cases{\left\{\,\vcenter\bgroup\normalbaselines\m@th \let\\=\@eqnacr
    \ialign\bgroup$##\hfil$&\quad##\hfil\crcr}
\def\endcases{\crcr\egroup\egroup\right.}

\def\eqalignno{\stepcounter{equation}\let\@currentlabel=\theequation
\if@defeqnsw\global\@eqnswtrue\else\global\@eqnswfalse\fi
\let\\=\@eqncr
$$\displ@y \tabskip\@centering \halign to \displaywidth\bgroup
  \global\@eqcnt\@ne\hfil
  $\@lign\displaystyle{##}$\tabskip\z@skip&\global\@eqcnt\tw@
  $\@lign\displaystyle{{}##}$\hfil\tabskip\@centering&
  \llap{\@lign##}\tabskip\z@skip\crcr}

\def\endeqalignno{\@@eqncr\egroup
      \global\advance\c@equation\m@ne$$\global\@ignoretrue}

\@namedef{eqalignno*}{\@defeqnswfalse\eqalignno}
\@namedef{endeqalignno*}{\endeqalignno}

\def\eqaligntwo{\stepcounter{equation}\let\@currentlabel=\theequation
\if@defeqnsw\global\@eqnswtrue\else\global\@eqnswfalse\fi
\let\\=\@eqncr
$$\displ@y \tabskip\@centering \halign to \displaywidth\bgroup
  \global\@eqcnt\m@ne\hfil
  $\@lign\displaystyle{##}$\tabskip\z@skip&\global\@eqcnt\z@
  $\@lign\displaystyle{{}##}$\hfil\qquad&\global\@eqcnt\@ne
  \hfil$\@lign\displaystyle{##}$&\global\@eqcnt\tw@
  $\@lign\displaystyle{{}##}$\hfil\tabskip\@centering&
  \llap{\@lign##}\tabskip\z@skip\crcr}

\def\endeqaligntwo{\@@eqncr\egroup
      \global\advance\c@equation\m@ne$$\global\@ignoretrue}

\@namedef{eqaligntwo*}{\@defeqnswfalse\eqaligntwo}
\@namedef{endeqaligntwo*}{\endeqaligntwo}

\newtoks\@stequation

\def\subequations{\refstepcounter{equation}%
  \edef\@savedequation{\the\c@equation}%
  \@stequation=\expandafter{\theequation}
  \edef\@savedtheequation{\the\@stequation}
  \edef\oldtheequation{\theequation}%
  \setcounter{equation}{0}%
  \def\theequation{\oldtheequation\alph{equation}}}

\def\endsubequations{%
  \setcounter{equation}{\@savedequation}%
  \@stequation=\expandafter{\@savedtheequation}%
  \edef\theequation{\the\@stequation}%
  \global\@ignoretrue}

\def\big#1{{\hbox{$\left#1\vcenter to1.428\ht\strutbox{}\right.\n@space$}}}
\def\Big#1{{\hbox{$\left#1\vcenter to2.142\ht\strutbox{}\right.\n@space$}}}
\def\bigg#1{{\hbox{$\left#1\vcenter to2.857\ht\strutbox{}\right.\n@space$}}}
\def\Bigg#1{{\hbox{$\left#1\vcenter to3.571\ht\strutbox{}\right.\n@space$}}}

\catcode`@=12

\begin{document}
\begin{titlepage}
\maketitle

\shorttitle{Metal--Insulator Transition}


\jnl{Journal of Physics: Condensed Matter}

\begin{abstract}
The critical exponents of the metal--insulator transition in disordered
systems have been the subject of much published work containing often
contradictory results.   Values ranging between $\half$ and $2$ can be
found even in the recent literature.   In this paper the results of a
long term study of the transition are presented.   The data have been
calculated with sufficient accuracy (0.2\%) that the calculated
exponent can be quoted as $s=\nu=1.54 \pm 0.08$ with confidence.   The
reasons for the previous scatter of results is discussed.
\end{abstract}
\thispagestyle{empty}
\vfill
\end{titlepage}

\section{Introduction}
The metal--insulator transition in disordered systems has been the
subject of theoretical and experimental work at least since Anderson
\shortcite{And58}.  The similarities with thermodynamic phase
transitions had been noted by several authors\cite{Tho74,Weg76} but it
was not until 1979 that a usable formulation of the renormalisation
group or scaling theory became available \cite{AALR79,Weg79c,Efe83}.
The basic assumption of these theories, that the behaviour could be
described by a single parameter scaling theory, was confirmed in
numerical calculations by the present author \cite{mKK81,mKK83a}.   For
a recent review of the area see Kramer and MacKinnon
\shortcite{KmK94}.

In spite of the progress made the exponents, $s$ and $\nu$, describing
the behaviour of the conductivity and the localisation length
respectively have proven difficult to calculate reliably.   For some
time there appeared to be a consensus between theory and experiment
that both exponents were equal to unity, but more recently this has
been called into question from both the theoretical (e.g.
\cite{KL84,Ler91a} ) and from  the experimental \cite{SHLMvH93}
side.

Numerical results have been scattered at least between $0.5$ and $2$
with numerous attempts at developing alternative methods of
calculation.   A good example of the difficulties is given by the
contrast between calculations for the Anderson model with rectangular
or Gaussian disorder\cite{KBmKS90}.   Using identical methods the
exponents obtained were about $1.5$ and $1.0$ for the rectangular and
Gaussian distributions respectively.   It is clearly unreasonable for
the exponents for these two cases to be different.   In fact if they
were different then it would call into question the justification of
the use of any simple model Hamiltonian to describe the transition and
so undermine the whole foundation of the subject.

In this paper the results of calculations carried out over several
years are presented.  All the basic results have an accuracy of at
least $0.2\%$ which enables the critical exponents to be calculated
much more accurately than when the conventional $1\%$ is used.

\section{Transfer Matrix Calculations}
The transfer matrix method has been discussed in numerous papers
\cite{mKK83a,PS81a} so only the briefest outline will be attempted
here.

The starting point is the usual Anderson\shortcite{And58} Hamiltonian
\be
H = \sum_i \epsilon_i |i><i| + \sum_{i\not=j} V_{ij}|i><j|
\ee{eq:1}
where $V_{ij} = V_0$ between nearest neighbours on a simple cubic
lattice and zero otherwise.  In this work $V_0 =1$ is chosen and will
therefore not be mentioned explicitly.  The diagonal elements
$\epsilon_i$ are independent random numbers chosen either from a
uniform rectangular distribution with $-\half W < \epsilon_i < +\half
W$ or from a Gaussian distribution of standard deviation $\sigma$.  For
purposes of comparison between the two cases an effective $W$ for the
Gaussian case may be defined by equating the variances as $W^2 =
12\sigma^2$ .

In terms of the coefficients $a_i$ of the wavefunctions on each site the
Schr\"odinger equation may be written in the form
\be
E a_i = \epsilon_i a_i + \sum_{j\not= i} a_j.
\ee{eq:2}

Consider now a long bar composed of $L$ slices of cross--section
$M\times M$.  By combining the $a_i$s from each slice into a vector
$\bi A_i$ (\ref{eq:2}) can be written in the concise form
\be
E \bi A_n = \bss H_n \bi A_n + \bi A_{n+1} + \bi A_{n-1}
\ee{eq:3}
where the subscripts $n$ now refer to slices and matrix $\bss H_n$ is
the Hamiltonian for slice $n$.    By rearranging (\ref{eq:3}) the
transfer matrix is obtained
\bes
\left(\ba{l}\bi A_{n+1}\\ \bi A_n\ea\right)
&=& \left(\ba{ll}E - \bss H_n&-\bss I\\
                \bss I                & 0\ea\right)
        \left(\ba{l} \bi A_n\\ \bi A_{n-1}\ea\right)\\
&=& \prod_{m=1}^{n} \left(\ba{ll}E - \bss H_m&-\bss I\\
                 \bss I                & 0\ea\right)
        \left(\ba{l} \bi A_1\\ \bi A_0\ea\right)\\
&=& \bss T_n \left(\ba{l} \bi A_1\\ \bi A_0\ea\right).
\ees{eq:4}
A theorem attributed to Oseledec\shortcite{Osc68} states that
\be
\lim_{n\to\infty} \left(\bss T_n^\dagger \bss T_n\right)^{1/n} = \bss M
\ee{eq:5}
where $\bss M$ is a well defined matrix and $\bss T_n$ are products of
random matrices.   The logarithms of the eigenvalues of $\bss M$ are
referred to as Lyapunov exponents and occur in pairs which are
reciprocals of one another.   By comparison with (\ref{eq:4}) the
Lyapunov exponents may be identified with the rate of exponential rise
(or fall) of the wave functions.  In fact the smallest exponent
corresponds to the longest decay length and hence to the localisation
length of the system.

In principle then it is necessary to calculate $\bss T_n$ for large
$n$, and diagonalise $\bss T^\dagger \bss T$.   Unfortunately the
calculation is not quite so simple: the different eigenvalues of $\bss
T^\dagger\bss T$ rise at different rates so that the smallest, which we
seek, rapidly becomes insignificant compared to the largest and is lost
in the numerical rounding error.  Typically this happens after about 10
steps.

\subsection{Orthogonalisation}
In order to obtain the smallest Lyapunov exponent it is necessary to
overcome this loss of numerical significance.  This can be achieved in
more than one way of which the orthogonalisation method is employed here.

After about 10 matrices have been multiplied together the columns of
the product matrix are orthogonalised to each other and normalised.
This is equivalent to multiplying the product from the right by an
appropriate matrix.   This orthonormalisation process automatically
separates the different exponentially growing contributions.

The process is repeated every 10 or so steps and the logarithm of the
length of the vector closest to unity is stored.  The Lyapunov exponent
is given by the mean value of these logarithms divided by the number of
steps between orthonormalisations.  In practice it is necessary to use
only 50\% or $M\times M$ vectors rather than the full $2\times M\times
M$ as the required vector is invariably the $M\times M$th.

The error in the Lyapunov exponent can be estimated from the variance
corresponding to the mean exponent.   Although this estimate could be
biased by correlations between the different contributions this is not
found to be a serious problem in practice, at least when the
localisation length is short compared with the distance between
orthogonalisation steps.

The optimum frequency of orthogonalisation steps can be estimated by
comparing the length of the $M\times M$th vector before and after
orthogonalisation.  The ratio should not be allowed to get close to the
machine accuracy.

\section{Scaling Theory}
The inverse of the smallest Lyapunov exponent is the localisation
length $\lambda_M$.   The renormalised length $\Lambda = \lambda_M/M$ is
found to obey a scaling theory\cite{mKK81,mKK83a} such that
\be
{\d\ln\Lambda\over\d\ln M} = {\mathop\chi\nolimits}\left(\ln\Lambda\right)
\ee{eq:6}
which has solutions of the form
\be
\Lambda = {\mathop{\rm f}\nolimits}\left(M/\xi\right)
\ee{eq:7}
where $\xi$ is a characteristic length scale which can be identified
with the localisation length of the insulator and which scales as the
reciprocal of the resistivity of the metallic phase\cite{mKK83a}.

In 3D (\ref{eq:6}) always has a fixed point $\chi = 0$ which corresponds
to the metal--insulator transition.   The behaviour close to the
transition can be found by linearising (\ref{eq:6}) and solving to
obtain
\be
\ln\Lambda = \ln\Lambda_c + A(\tau - \tau_c)M^\alpha
\ee{eq:8}
where $\tau$ is the disorder $W$ or $\sigma$, $\Lambda_c$ and $\tau_c$
represent the critical $\Lambda$ and disorder respectively, and $A$ and
$\alpha$ are constants.  By comparing (\ref{eq:7}) and (\ref{eq:8}) an
expression for $\xi$ can be obtained in the form
\be
\xi \sim \left|\tau - \tau_c\right|^{1/\alpha}
\ee{eq:9}
so that the localisation length exponent $\nu$ is given by $\nu =
1/\alpha$.   Since it is well known\cite{Weg76,AALR79} that the
conductivity exponent $s$ is related to $\nu$ by $s = (d-2)\nu$ then by
fitting (\ref{eq:8}) to the data and calculating $\alpha$ both exponents
can be obtained.

\subsection{Deviations from Scaling}
One simple feature of (\ref{eq:8}) is that, when $\ln\Lambda$ is
plotted against $\tau$, the curves for different $M$ intersect at a
common point $(\ln\Lambda_c, \tau_c)$.   In practice the data do not
behave in exactly this way.  There is a small deviation from scaling.
This deviation could be taken into account by adding an extra term to
(\ref{eq:8}) which depends on $M$ but not on $\tau$.  Consider, however,
the form
\be
\ln\Lambda = A\tau M^\alpha + B(M)
\ee{eq:10}
which represents the most general form of such a correction.  If a
specific form for the correction were assumed it would require at least
4 independent fitting parameters to represent $B(M)$, including
$\Lambda_c$ and $\tau_c$, and may still not represent the true deviation
from scaling.  It seems better therefore to fit an independent $B(M)$
for each value of $M$ and therefore to make no assumption about the
nature of the deviation from scaling, other than that it is
non--critical, and therefore independent of $\tau$, in the region of
interest.  By fitting the data to (\ref{eq:10}) in this way the
exponent $\alpha$ is derived solely from the gradient of $\ln\Lambda$
\mbox{vs.} $\tau$ and the intercept is allowed to float.  The results of
such fits are shown in figure~\ref{fig:1}.

\subsection{Data Fitting}
The data can be fitted to (\ref{eq:10}) by iteratively using a standard
least squares procedure.  Care is required with the non--linear parameter
$\alpha$.  The quality of the fit can be tested by computing $\chi^2$
defined as
\be
\chi^2 = \sum_i{\left(A\tau_i M_i^\alpha + B(M_i) -
\ln\Lambda_i\right)^2
\over\sigma^2_i}
\ee{eq:11}
where $i$ runs over all data points and $\sigma_i$ is the error in
point $i$.  After fitting $\chi^2$ should be approximately equal to the
number of data points less the number of fitted parameters.   Hence the
value of $\chi^2$ provides a measure of the quality of the fit.   In
the results presented here the range of values of disorder round the
critical value was chosen such that $\chi^2$ conforms to this
condition.  Then a large number of additional points was calculated
inside this range.  An important side effect of this procedure is that
the apparently acceptable range of disorder around the fixed point gets
narrower
as the calculations become more accurate.  It is therefore important to
test whether any apparent change in the fitted exponent is due to this
narrowing.

The values of the ideal and the fitted $\chi^2$ as well as the range
considered are shown in table~\ref{tab:1}. Using $4\le M\le 12$ and the
widest range of disorder $s=\nu=1.53\pm 0.04$ and $s=\nu=1.48\pm 0.05$
for rectangular and Gaussian cases respectively.

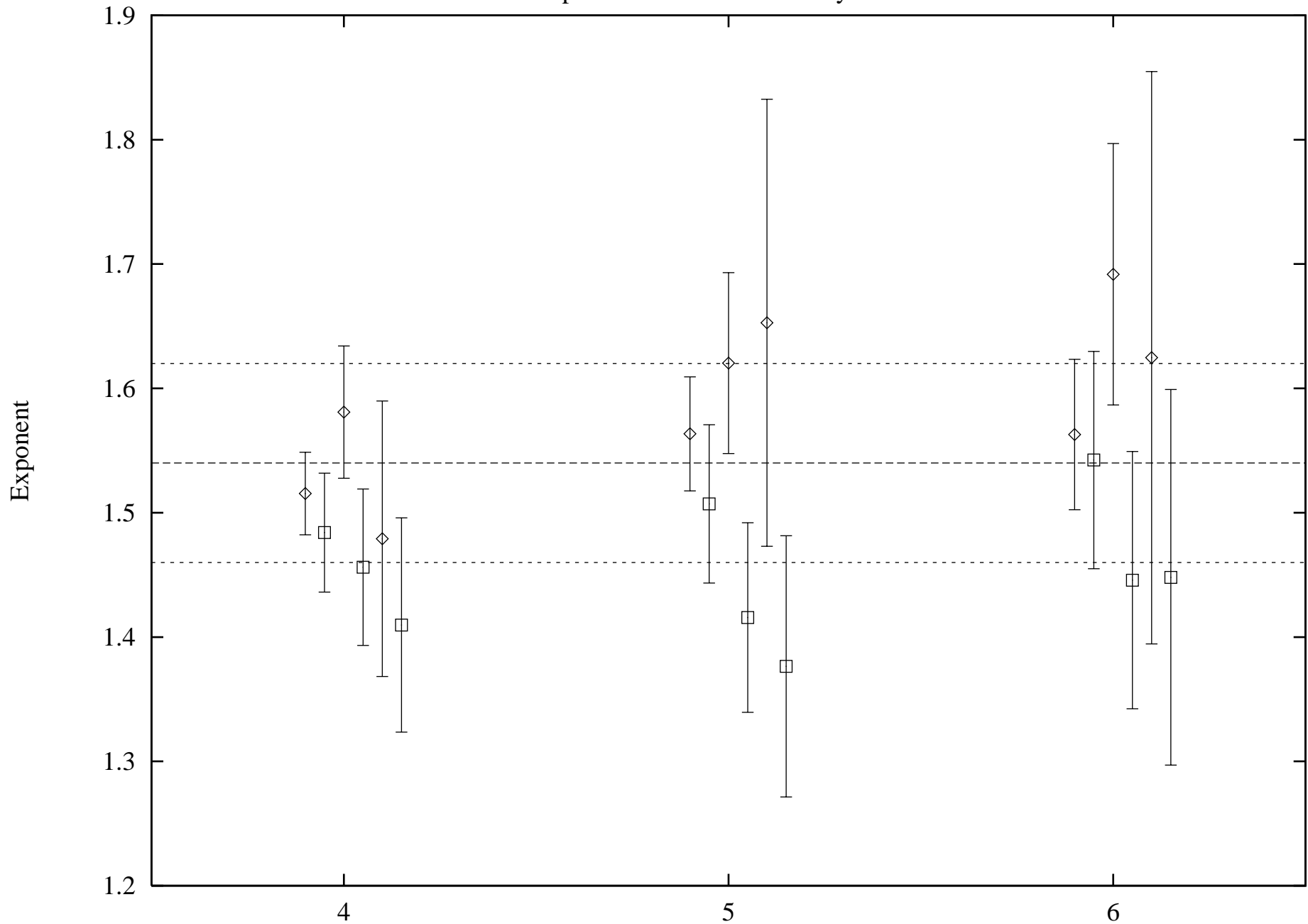
\subsection{Statistical and Systematic Errors}
The statistical error in the fitted critical exponent is easily
estimated from the least squares fitting procedure.  Systematic errors
are more difficult to take into account.  In this work an attempt is
made to consider 3 sources of systematic error:
\begin{itemize}
\item Limited range of system sizes:  $4\le M\le 12$ has been considered
and the effect of ignoring the smaller system sizes tested.
\item Width of the critical region: the maximum range of disorder is
imposed by $\chi^2$ but may still be too large.  The effect of narrowing
this range still further has been tested.
\item The choice of distribution of random numbers: this has been
tested by comparing the rectangular and Gaussian cases.
\end{itemize}
These tests are represented in figure~\ref{fig:2}.  Unfortunately the
general increase in the error bars due to ignoring data tends to mask
any systematic changes.  There does however appear to be a general
increase in the exponents when the $M=4$ data is eliminated and a
tendency for the Gaussian data to lie below the rectangular.  From this
data $s=\nu\approx 1.54\pm0.08$ has been estimated, where the error bar
may be somewhat wider than necessary.

\section{Results and Conclusions}
The results are summarised in table~\ref{tab:1}.  All these results
have been calculated in the middle of the band (i.e. $E=0$), but there
is ample evidence that for the models considered here, this point is
not special and is truly representative of the whole band, at least in the
range $-6<E<6$.
\begin{table}[htbp]
\begin{tabular*}{\textwidth}{|l@{\extracolsep{\fill}}c|cc|cc|} \hline
			&&\bf Rectangular	&&\bf Gaussian	&\\ \hline
Exponent		&&$1.515\pm 0.033$	&&$1.484\pm0.048$&\\
Disorder Range		&&$16.2\le W\le 16.8$	&&$21.0\le W\le21.5$&\\
System Sizes		&&$4\le M\le 12$	&&$4\le M\le 12$&\\
$\chi^2$(expected) 	&&$142$			&&$97$		&\\
$\chi^2$(fitted)	&&$126$			&&$75$		&\\
$W_c$			&&$16.50\pm 0.05$	&&$21.20\pm0.06$&\\
$\sigma_c$		&&$4.763\pm 0.015$	&&$6.120\pm0.018$&\\
$\Lambda_c$		&&$0.580\pm0.005$	&&$0.580\pm 0.005$&\\ \hline
\end{tabular*}
\caption{\label{tab:1}
N.B: The estimates of $W_c$ and $\Lambda_c$ are based on the values
given by several different fitting procedures.}
\end{table}

Unlike previous calculations \cite{KBmKS90} the exponents calculated
for the two distributions now overlap well and are therefore consistent
with the common assumption that simply changing the distribution does
not change the universality class and hence the critical exponent.  The
discrepancy reported previously is presumably due to insufficient
accuracy in the raw data and consequent assumption of a critical range
of disorder which was too wide.

This may have consequences for experiment as it seems to suggest that
it is possible to obtain an exponent of unity simply by using too wide
a range of data around the critical disorder, energy, pressure, etc.  It
should also be borne in mind that the influence of interactions may also
account for differences between experimental results and those based on
a model of non--interacting electrons.  For this reason it may be more
realistic to compare the present results with photonic or acoustic
rather than electronic experiments.

In summary, the critical exponent of the Anderson model of the
metal--insulator transition is $s=\nu=1.54\pm 0.08$.

\section*{Acknowledgements} This work has profited from many useful
discussions with B.Kramer, M.Schreiber, J.B.Pendry, P.M.Bell,
R.B.S.Oakeshott, E.A.Johnson, and P.J.Roberts.  The financial support
of the UK SERC and the European Union, through SCIENCE grant
$\mbox{SCC}^*$--CT90--0020, is gratefully acknowledged.

\section*{Figure Captions}
\begin{enumerate}
\item\label{fig:1} $\Lambda$ \mbox{vs.} $W$, for (a)
rectangular and (b) Gaussian distributions. The data are represented
by dots with differing symbols for different system sizes with $4\le
M\le 12$ increasing in the direction of the arrow.  Each point is
accurate to $0.2\%$.  The lines are fitted using (\ref{eq:10}).

\item\label{fig:2}  Fitted critical exponents for rectangular
(Diamonds) and Gaussian (Squares) distributions.  The absciss{\ae}
represent the smallest system size taken into account (with small
offsets for clarity).  In each group the width of the fitted region is
(from left to right)
$(16.2 \le W \le 16.8)\to(16.3\le W\le 16.7)\to(16.4\le W\le 16.6)$ and
$(21.0\le W\le 21.5)\to(21.05\le W\le 21.45)\to(21.1\le W\le 21.4)$ for
rectangular and Gaussian cases respectively.  The dotted lines represent
the range  $s=\nu=1.54\pm 0.8$.
\end{enumerate}


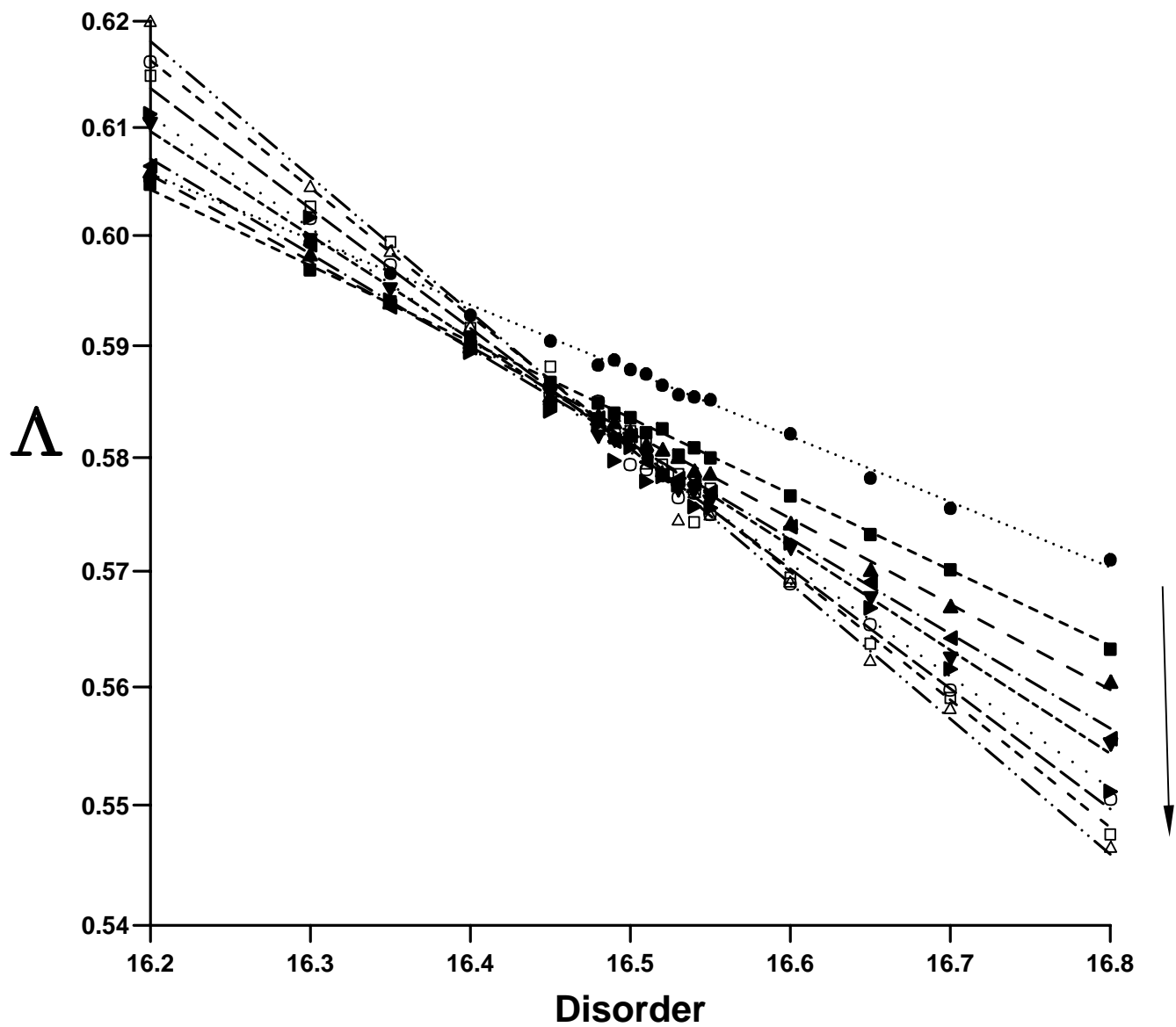
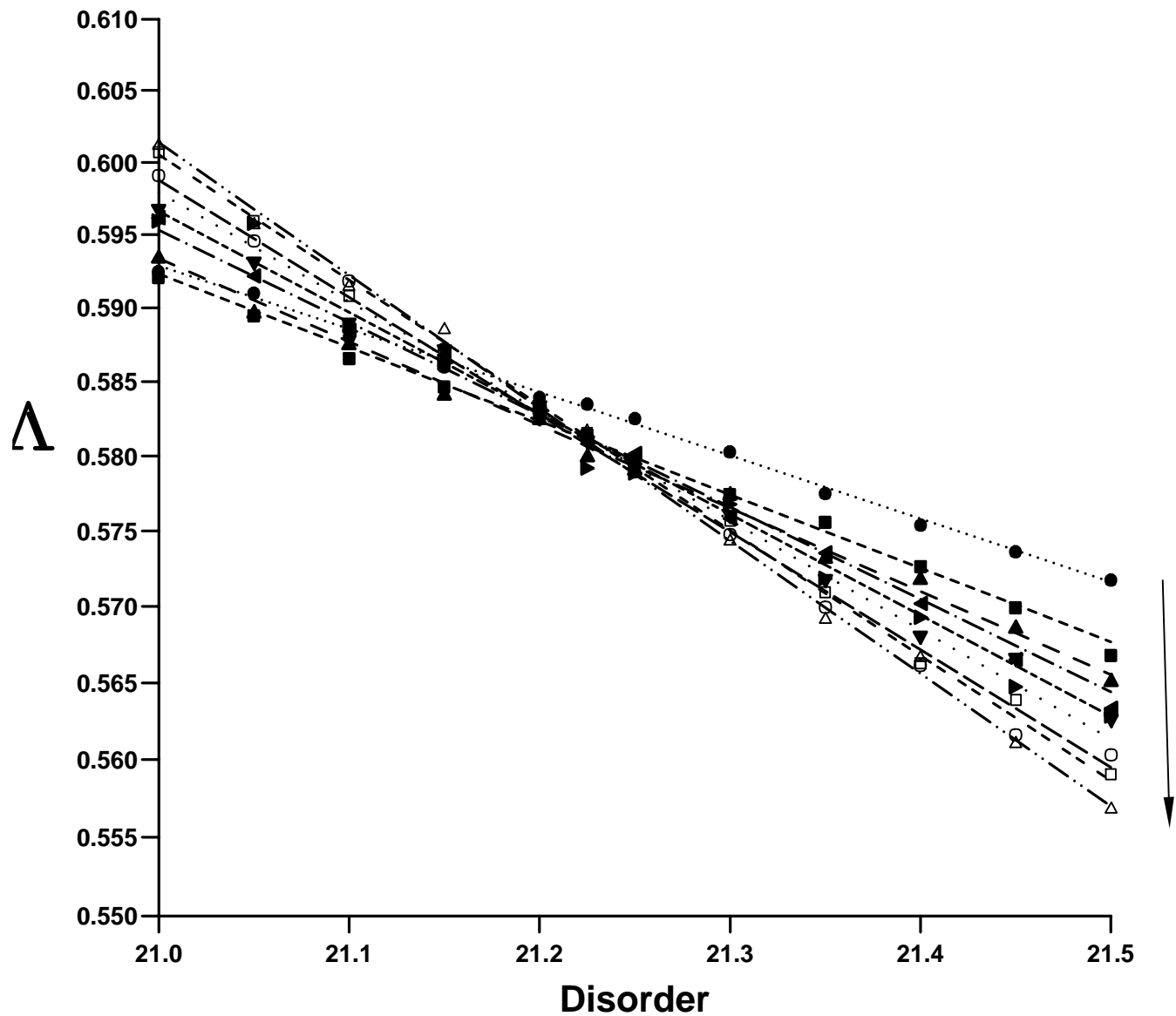
\end{document}